\documentclass[letterpaper]{article} 
\usepackage{aaai2026}  
\usepackage{times}  
\usepackage{helvet}  
\usepackage{courier}  
\usepackage[hyphens]{url}  
\usepackage{graphicx} 
\usepackage{natbib}  
\usepackage{caption} 
\usepackage{algorithm}
\usepackage{algorithmic}

\usepackage{newfloat}
\usepackage{listings}
\DeclareCaptionStyle{ruled}{labelfont=normalfont,labelsep=colon,strut=off} 
\floatstyle{ruled}
\newfloat{listing}{tb}{lst}{}
\floatname{listing}{Listing}
\newcommand{\workname}{MTMC}
\usepackage{xcolor}
\usepackage{booktabs}
\usepackage[table]{xcolor}
\usepackage{multirow}
\usepackage{amsmath}
\usepackage{makecell}

\title{QiMeng-Kernel: Macro-Thinking Micro-Coding Paradigm for LLM-Based High-Performance GPU Kernel Generation}
\author{
    Xinguo Zhu\textsuperscript{\rm 1,\rm 3, \rm 4},
    Shaohui Peng\textsuperscript{\rm 1, \rm 4}\thanks{Corresponding author.},
    Jiaming Guo\textsuperscript{\rm 2},
    Yunji Chen\textsuperscript{\rm 2, \rm 4},
    Qi Guo\textsuperscript{\rm 2},
    Yuanbo Wen\textsuperscript{\rm 2},
    Hang Qin\textsuperscript{\rm 1,\rm 3, \rm 4},
    Ruizhi Chen\textsuperscript{\rm 1, \rm 4},
    Qirui Zhou\textsuperscript{\rm 2, \rm 4},
    Ke Gao\textsuperscript{\rm 1, \rm 4},
    Yanjun Wu\textsuperscript{\rm 1, \rm 4},
    Chen Zhao\textsuperscript{\rm 1, \rm 4},
    Ling Li\textsuperscript{\rm 1, \rm 4}\footnotemark[1]
}
\affiliations{
    \textsuperscript{\rm 1} Intelligent Software Research Center, Institute of Software, CAS, Beijing, China
    \\
    \textsuperscript{\rm 2} State Key Lab of Processors, Institute of Computing Technology, CAS, Beijing, China
    \\
    \textsuperscript{\rm 3} Hangzhou Institute for Advanced Study, UCAS, Hangzhou, China
    \\
    \textsuperscript{\rm 4} University of Chinese Academy of Sciences

    zhuxinguo23@mails.ucas.ac.cn, pengshaohui@iscas.ac.cn, liling@iscas.ac.cn
}

\usepackage{bibentry}

\begin{document}

\maketitle

\begin{abstract}
Developing high-performance GPU kernels is critical for AI and scientific computing, but remains challenging due to its reliance on expert crafting and poor portability.
While large language models (LLMs) offer promise for automation, both general-purpose and finetuned LLMs suffer from two fundamental and conflicting limitations: correctness and efficiency.
The key reason is that existing LLM-based approaches directly generate the entire optimized low-level programs, requiring exploration of an extremely vast space encompassing both optimization policies and implementation codes.

To address the challenge of exploring an intractable space, 
we propose Macro Thinking Micro Coding (\workname), a hierarchical framework inspired by the staged optimization strategy of human experts. It decouples optimization strategy from implementation details, ensuring  efficiency through high-level strategy and correctness through low-level implementation.
Specifically, Macro Thinking employs reinforcement learning to guide lightweight LLMs in efficiently exploring and learning semantic optimization strategies that maximize hardware utilization.
Micro Coding leverages general-purpose LLMs to incrementally implement the stepwise optimization proposals from Macro Thinking, avoiding full-kernel generation errors.
Together, they effectively navigate the vast optimization space and intricate implementation details, enabling LLMs for high-performance GPU kernel generation.

Comprehensive results on widely adopted benchmarks demonstrate the superior performance of {\workname} on GPU kernel generation in both accuracy and running time.
On KernelBench, MTMC achieves near 100\% and 70\% accuracy at Levels 1-2 and 3, over 50\% than SOTA general-purpose and domain-finetuned LLMs, with up to 7.3$\times$ speedup over LLMs, and 2.2$\times$ over expert-optimized PyTorch Eager kernels. On the more challenging TritonBench, MTMC attains up to 59.64\% accuracy and 34$\times$ speedup. All models and datasets will be made publicly available.
\end{abstract}


\section{Introduction}
As the cornerstone of modern artificial intelligence (AI) computing, scientific simulation, and large-scale analytics, GPU kernels face growing demand for high performance, fueled by the quest for computational efficiency across diverse hardware and algorithms \cite{paszke2019pytorch, Efficient-Transformers-survey, pandey2022transformational, ansel2024pytorch}. 
\begin{figure}[t]
    \centering
    \includegraphics[width=1.0\columnwidth]{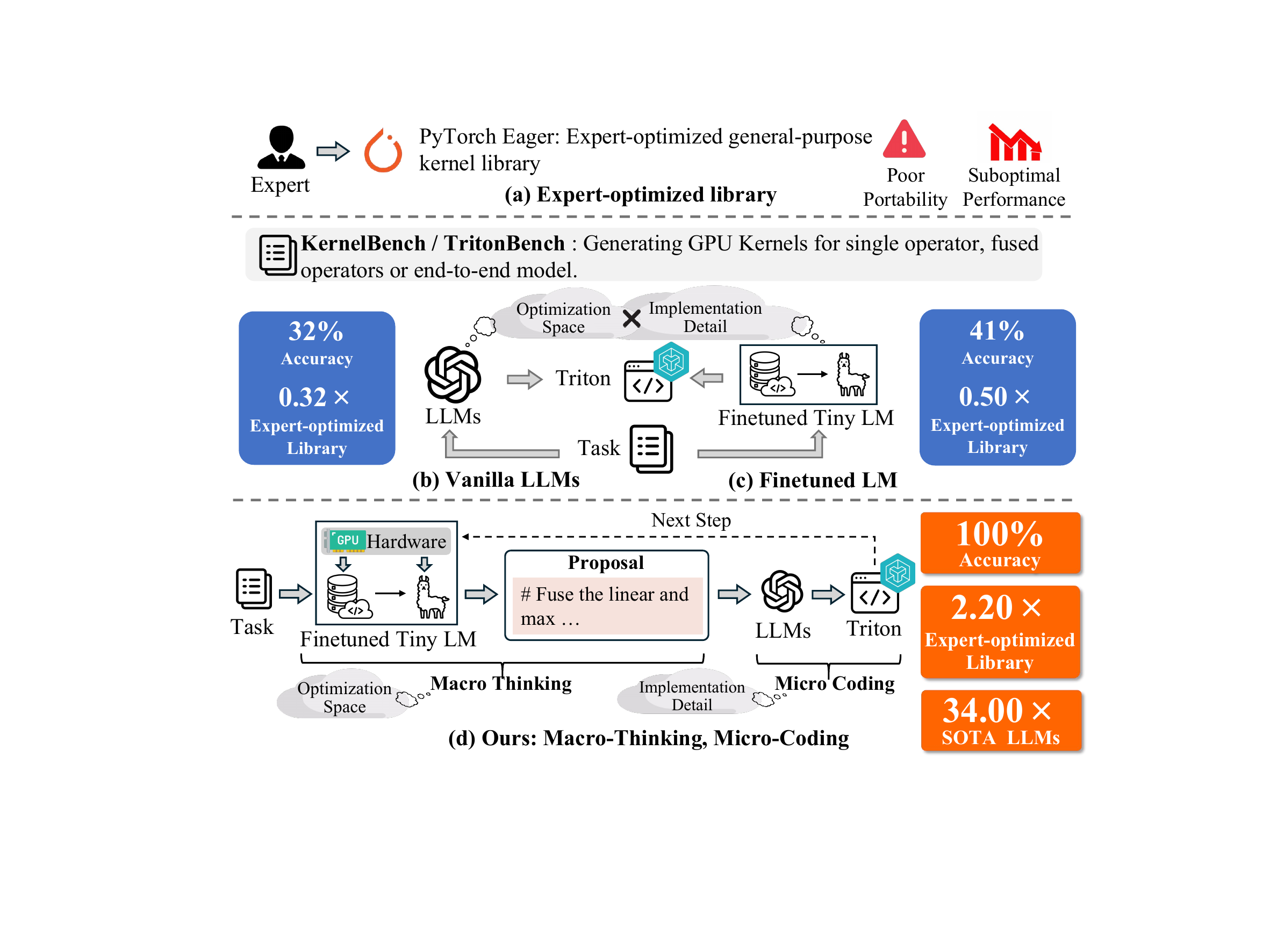}
    \caption{Comparison of GPU kernel generation paradigms.}
    \label{fig:intro}
\end{figure}
Yet developing high-performance GPU kernels remains notoriously challenging, heavily reliant on experts’ empirical knowledge and hardware insight for the manual tuning process, which is prohibitively expensive and inefficient.
For example, FlashAttention implementation on Hopper architecture takes years-long development \cite{dao2022flashattention, dao2023flashattention, shah2024flashattention} with poor portability.
Despite the emergence of GPU Domain-Specific Language (DSL) like Triton \cite{tillet2019triton}, experts remain essential for designing hardware-specific optimization strategies and data layouts, failing to resolve the inherent complexities or platform portability issues.
While established expert-optimized libraries (e.g., generic kernels in PyTorch Eager, as shown in Figure \ref{fig:intro}(a)) often sacrifice peak performance for specific scenarios, and suffer from limited cross-hardware portability.
Consequently, automated generation of high-performance GPU kernels has emerged as a critical and persistent challenge.

Given the rapid advances of LLMs in code generation tasks \cite{le2022coderl, joshi2023repair, jiang2024survey, tian2025codehalu}, a growing body of research is now focusing on leveraging LLMs to automatically generate high-performance GPU kernels \cite{qimeng:tensorop, qimeng:gemm}. 
Existing methods employ either powerful general-purpose LLMs or finetuned specialized LLMs, but neither delivers satisfactory performance, as shown in Figure \ref{fig:intro}(b) and (c).
Recent most widely adopted benchmarks (KernelBench \cite{ouyang2025kernelbench}, TritonBench\cite{li2025tritonbench}) show that even SOTA general-purpose LLMs struggle to generate correct GPU kernels.
They produce significant errors ranging from basic compilation failures to computational flaws across both CUDA and Triton implementations, while performing substantially worse than expert-crafted kernels.
Due to the extremely limited availability of optimized kernel datasets \cite{li2023starcoder}, domain-specific finetuned LLMs (e.g., KernelLLM developed by Meta \cite{kernelllm2025}, Kevin-32B from Stanford  \cite{multi-turn-kernels}) also suffer from inadequate performance and cross-task generalization.
Overall, while LLM-based methods represent a promising direction for automating high-performance GPU kernel generation, they suffer from two fundamental and conflicting limitations: correctness and efficiency.

The core challenge stems from the extreme complexity inherent in the GPU kernel optimization space (already $\approx 10^9$ for one subgraph of neural networks on GPU \cite{zhai2024enabling}) coupled with implementation details (hundreds/thousands of code lines) under low-level hardware-specific constraints, which is fundamentally distinct from traditional code generation tasks.
Without precise hardware and optimization understanding, LLMs prove incapable of navigating the vast optimization space or managing intricate implementation details, as shown by analysis in KernelBench.
High-performance kernels require sophisticated pipeline parallelization and multi-level memory access patterns tailored to specific hardware, necessitating extensive tuning even by experienced experts. 
Moreover, their complex implementations are vulnerable to minor errors that propagate performance inefficiencies or faults.
Thus, prior LLM-based approaches are fundamentally inadequate, generating kernels that are either incorrect or significantly underoptimized.

To address the above core challenge, we innovatively decouple GPU kernel generation into high-level optimization strategy  and stepwise optimized code implementation, targeting to enable LLM-based generation of correct, high-performance kernel code with minimal data tuning and human effort, as shown in Figure \ref{fig:intro}(d).
Based on this insight, we propose a novel hierarchical GPU kernel generation framework, Macro Thinking Micro Coding (MTMC).
The high-level Macro Thinking progressively generates semantic optimization proposals based on hardware characteristics without implementation details.
We utilize reinforcement learning (RL) to drive lightweight LLMs in efficiently learning optimization policies on a compact human-curated dataset to maximize hardware utilization. 
Simultaneously, the lower-level Micro Coding leverages general-purpose LLMs for stepwise optimization implementation, circumventing errors inherent in whole kernel generation.
This decoupled paradigm enables Macro Thinking to eliminate massive finetuning data requirements while empowering Micro Coding to harness LLMs' proficiency in atomic code implementation.
Thus, MTMC facilitates the automated generation of correct and high-performance kernels.

The key contributions of this paper are as follows:

\begin{itemize}
\item We introduce a novel hierarchical generation paradigm that effectively decouples complex optimization policy design from low-level implementation, enabling LLM to generate high-performance GPU kernels.

\item We construct a compact yet effective dataset and environment for efficient kernel optimization policy training, which enables step-by-step guidance for general-purpose LLMs to generate high-performance GPU kernels.

\item Results on widely adopted benchmarks show {\workname} achieves outstanding performance of GPU kernel generation in both accuracy and running time.
On KernelBench, MTMC achieves accuracy of near 100\% and 70\% at Levels 1-2 and 3 respectively (over 50\% than general-purpose and domain-finetuned LLMs) with up to 7.3$\times$ speedup over LLMs, and 2.2$\times$ over expert-optimized PyTorch Eager kernels. Furthermore, MTMC attains remarkable gains on the more challenging TritonBench, up to 59.64\% accuracy and 34 $\times$ speedup.
\end{itemize}

\begin{figure*}[htbp]
    \centering
    \includegraphics[width=0.8\textwidth]{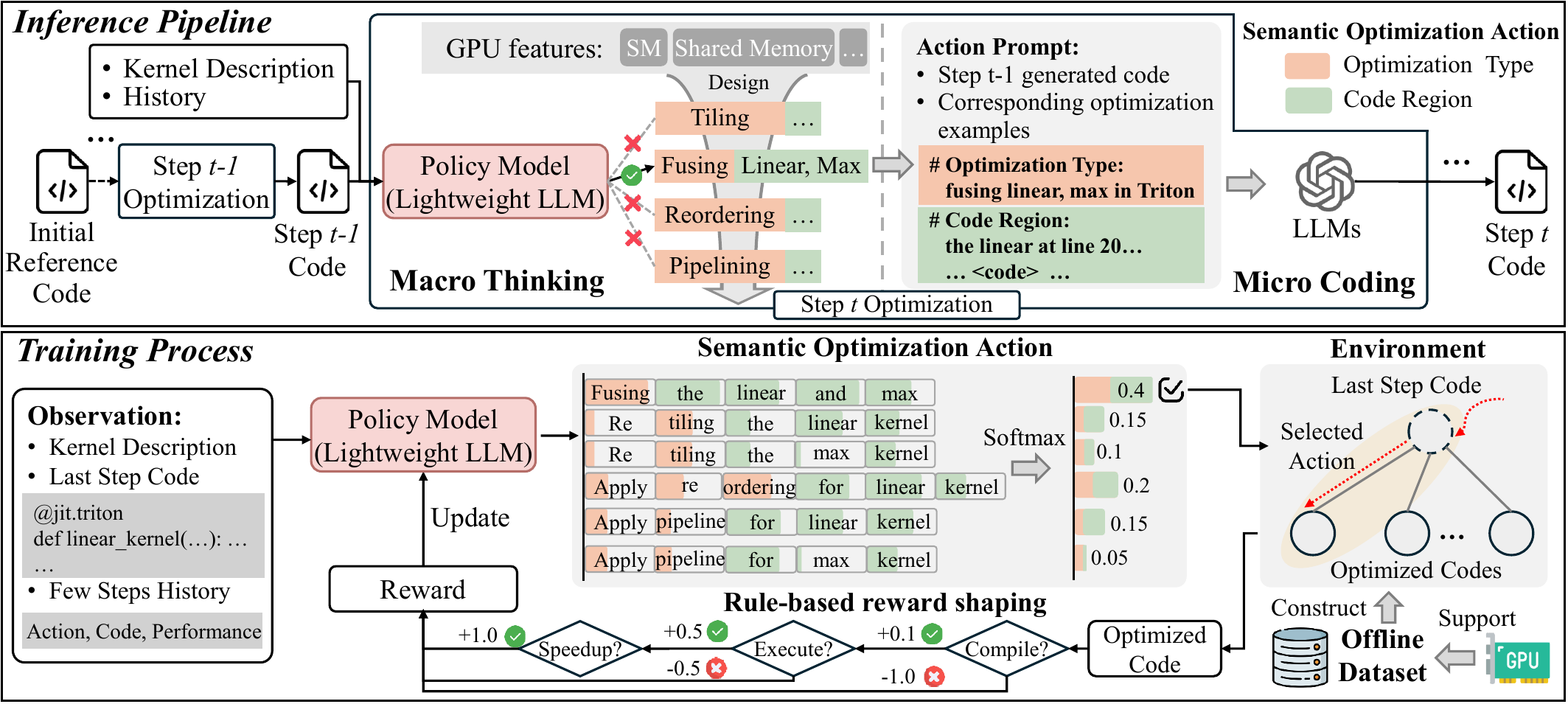}
    \caption{{\workname} overview. The framework takes unoptimized PyTorch code as input and generates high-performance GPU kernels with hierarchical process: Macro Thinking generates semantic optimization actions, while Micro Coding implements them step-by-step. The optimization policy based on lightweight LLMs is trained with RL on compact human-crafted dataset.
    }
    \label{fig:method}
\end{figure*}

\section{Related Works}

\paragraph{Benchmarks}

KernelBench \cite{ouyang2025kernelbench} and TritonBench \cite{li2025tritonbench} are widely adopted GPU kernel generation benchmarks to evaluate the capability of LLMs to generate efficient GPU kernels.

\begin{table}[htbp]
\centering
\scriptsize
\setlength{\tabcolsep}{2pt}
\begin{tabular}{l | l | l}
    \toprule
    \textbf{Benchmark} & \textbf{Description} & \textbf{Types} \\
    \midrule
    \multirow{3}{*}{\textbf{KernelBench}} 
    & Level 1: 100 single ops & GEMM, Convolution, Softmax, ... \\
    & Level 2: 100 fused ops & GEMM+Max, Conv2d+ReLU, ...\\
    & Level 3: 50 neural networks & LSTM, VGG16, MiniGPT, ViT,...\\
    \midrule
    \multirow{2}{*}{\textbf{TritonBench}} 
    & T: 166 common tasks &  Adam, SGD, BatchNorm, Argmax ...\\
    & G: 184 real-world cases & FlashAttention, BMM, Cumsum, ... \\
    \bottomrule
\end{tabular}
\caption{Details of KernelBench and TritonBench.}
\label{tab:kb_details}
\end{table}

\paragraph{LLM-based Kernel Generation}
The traditional LLM-based code generation focuses on generating functionally correct high-level programming language codes, and has seen rapid development \cite{sun2020treegen, shin2021survey, li2022competition, chen2024divide, jiang2024survey}.
General-purpose LLMs (like Gemini 2.5 Pro \cite{comanici2025gemini} and Claude Sonnet 4 \cite{claude4}) and code LLMs (like Qwen-Coder \cite{hui2024qwen2} and DeepSeek-Coder \cite{guo2024deepseek, zhu2024deepseek}) have demonstrated strong capabilities in understanding and generating high-level programming languages across diverse tasks.
However, these models struggle with generating correct and high-performance GPU kernels, as they lack comprehensive hardware and optimization understanding. 
Given the complexity and unique challenges of high-performance kernel generation, several targeted approaches have emerged. 
The QiMeng series  \cite{qimeng:attention, qimeng:gemm, qimeng:tensorop, dong2025qimeng} focuses on leveraging LLMs to generate specific operators through hardware-aware generation and parameter tuning.
KernelLLM and Kevin-32B leverage small sets of kernel data to finetune LLMs. 
Additionally, there are also agent-based approaches like AI CUDA Engineer.
However, these works are often constrained by specific operators and limited by data scarcity, resulting in poor generalization and suboptimal performance. 
Different from previous works, our {\workname} decouples high-level strategy from low-level details, achieving efficient learning of general optimization and high correctness.

\section{Preliminary}
\subsection{GPU Kernel Generation Task}
GPU kernel generation constitutes a specialized program synthesis task \cite{manna1971toward, waldinger1969prow}, aiming to produce correct and high-performance GPU kernel implementations $P$ from abstract algorithmic specifications $S$ (e.g., natural language descriptions, simple high-level reference code like Python) on a specific GPU $h$:
\begin{equation}
\small
\label{eq:f} 
\begin{array}{ll}
    \underset{P}{\textbf{minimize}} & \mathrm{RunningTime}(P, h) \\[2ex]
    \textbf{subject to} & P \equiv S  ~\& ~\mathrm{Correct}(P, h) = \mathrm{true}
\end{array}
\end{equation}

Compared to other code generation tasks, GPU kernel generation presents significantly greater challenges.
It necessitates simultaneously satisfying hardware constraints while achieving both functional correctness and high performance.
The former requires successful compilation and execution on target hardware with accurate computational outputs; the latter demands maximally leveraging GPUs' complex memory hierarchy and parallel computing units to attain high memory/compute efficiency.
This represents a fundamental trade-off: higher-performance optimizations typically entail more intricate implementation details, heightening error susceptibility.
Even experts must rely on profound hardware architecture and optimization comprehension, undergoing iterative refinement.

\subsection{Optimization Principles}\label{sec:optimization tech}
We summarize four fundamental optimization principles corresponding to specific aspects of GPU characteristics to transform a naive algorithm into a complex, high-performance GPU kernel implementation.
\begin{itemize}
\item \textbf{Tiling} partitions data to fit share memory size for memory access acceleration \cite{dally2021evolution}. 
\item \textbf{Fusion} merges operators to reduce memory access costs, especially in memory-bound workloads \cite{shi2023welder, snider2023operator, wu2025mirage, shah2024flashattention}.
\item \textbf{Pipeline} overlaps computation and data movement to improve hardware utilization efficiency \cite{tan2011fast}. 
\item \textbf{Reordering} swaps loops to promote memory access locality \cite{anam2013precision}. 
\end{itemize}

\section{Method}
In this section, we introduce the {\workname} framework, as in Figure \ref{fig:method}, which enables LLMs for high-performance GPU kernel generation by decoupling the optimization and implementation stages to address the efficiency and correctness challenges respectively.
The inference pipeline of {\workname} and the training process of optimization policy will be detailed in subsequent subsections.

\subsection{Inference Pipeline}
The Inference Pipeline consists of Macro Thinking and Micro Coding, which decouple the GPU kernel generation into multi-step high-level optimization and corresponding low-level code implementation.
Specifically, lightweight LLMs iteratively generate semantic optimization proposals based on a learned policy to guide general-purpose LLMs in step-by-step kernel generation, in order to ensure maximal correctness while maximizing hardware utilization.

\textbf{Macro Thinking.} 
It leverages the finetuned lightweight LLMs (training details in Section \ref{chap: training}) to iteratively generate semantic optimization actions based on hardware information and the code to be optimized. 
Macro Thinking formulates the optimization process as a stepwise decision-making problem.
Each step, the finetuned lightweight LLM policy takes last step kernel code, kernel description and history information as input, then outputs semantic optimization action to Micro Coding stage for implementation, until terminal action or max step is reached.
The action declares:
(1) the \emph{Optimization Type} (e.g., “fusing”), and
(2) the \emph{Code Region} indicating where the optimization should be applied.

Such design enables the policy to generate effective optimization proposals to guide lower-level implementation.
First, actions are carefully designed to exploit diverse aspects of GPU hardware features.
Second, semantic macro-actions enable efficient policy exploration and learning from offline data on leveraging GPU features performance optimization without concerning of implementation details.

\textbf{Micro Coding. } 
Micro Coding focuses on translating semantic optimization actions into syntactically correct optimized implementations on kernel.
Given the action prompt, Micro Coding output next step kernel code for new round of iteration, by leverages general-purpose LLMs to generate low-level code modifications that implement the intended optimization.
As illustrated in Figure \ref{fig:method}, the action prompt is automatically constructed by three elements: (1) kernel code of step $t-1$, (2) the semantic optimization action (including type and region) from Macro Thinking, and (3) examples for corresponding optimization type.

Precisely because Macro Thinking demands only atomic, single-step optimizations with explicitly specified type and region, Micro Coding can leverage in-context learning to maximize the probability of generating correct code modifications.
With iterative refinement, {\workname} ultimately produces correct and high-performance GPU kernels.

\subsection{Training Process}
\label{chap: training}
To efficiently explore the optimization space and learn an effective optimization policy, we finetune a lightweight LLMs via RL to exploit hardware features. 
In this section, we detail the training process, focusing on four key aspects: the semantic optimization action space, the training methodology, environment and reward design, as shown in Figure \ref{fig:method} (We provide more training details in the Appendix).

\textbf{Semantic Optimization Action Space.}
{\workname} enables the policy model to explore and learn kernel optimization strategies by semantically representing optimization actions.
Semantic optimization action space are combination of optimization types and the candidate code regions.
The optimization types are derived from experts' kernel optimization experience, which focus on leverage different aspects of GPU features.
The candidate code regions are determined based on the data flow and abstract syntax tree (AST) analysis to identify syntactically and semantically valid code segment.
For example, ``fusing the linear and max in line 15 to 20'' means fusing adjacent operators to reduce memory access.
In summary, the action space refines and extends optimization techniques introduced in Section \ref{sec:optimization tech}, including Tiling, Fusion, reordering, and pipeline.
Such action space design is representative for hardware exploitation and can narrow the optimization space, thus supporting Macro Thinking policy efficiently exploring and learning.
 
\textbf{Training Methodology.} 
The Macro Thinking policy is a lightweight pre-trained LLM (e.g. DeepSeek-Coder-1.3B and Llama-3.2-1B) , which takes observation as input and then produces semantical optimization actions, as shown in Figure \ref{fig:method}.
The semantic optimization action {\small $a_k \in \mathcal{A}$} is a sequence of tokens {\small$a_k = \{w_k^1, \dots, w_k^{N_k}\}$}, where {\small$N_k$} is the length of the semantic optimization action {\small$k$}.
The action sampling ratio equals to joint probability of all tokens: 
\begin{equation}
\small
\begin{split}
    P_{\mathrm{token}}(a_k | s) &= \prod_{i=1}^{N_k} P(w_k^i | s, w_k^1, \dots, w_k^{i-1})
\end{split}
\label{eq:norm}
\end{equation}

The sampling probability for each action is then derived by applying Softmax normalization to the logits.
We adopt TWOSOME training framework \cite{tan2024true} and the standard Proximal Policy Optimization (PPO) \cite{schulman2017proximal} algorithm to train the policy.

\textbf{Environment.}
For representative GPU kernel tasks, including distinct single operations, subgraphs and entire neural networks, we collect a representative \textbf{offline dataset} comprising 60k trajectories, without benchmark instances.
Each task contains optimization trajectories traversing multiple actions from initial reference code, which provides the foundation for Macro Thinking policy learning.
With these offline trajectories, we construct a tree-structured RL environment, as shown in Figure \ref{fig:method}.
When applying a selected semantic optimization action to current kernel code nodes, the state would transit to varying leaf nodes to provide next step kernel code.
This design primarily enables efficient policy training by avoiding prohibitive latency from real-time interaction with LLMs in Micro Coding stage.

\textbf{Reward Shaping.}
To guide the learning process, we adopt a rule-based reward shaping across the following criteria.
Rewards are assigned based on three criteria from easy to hard:
(1) successful compilation,
(2) correct executable results, running time without errors, and
(3) performance improvement over the previous kernel.
Rewards increase progressively while penalties decrease gradually, ensuring the policy initiates exploration from valid optimizations and ultimately learns highly optimized solutions.
We further set a step-proportional reward decay mechanism to mitigate degenerate looping behaviors during policy exploration.

\section{Results}
\subsection{Experiment Setup}
To study the performance and generality of {\workname}, comprehensive evaluations are conducted across 3 different hardware platforms, 13 distinct LLMs and agent, and 2 widely adopted benchmarks (some results refer to Appendix).

\textbf{LLMs.} 
Marco Thinking is validated with 3 lightweight LLMs, including DeepSeek-Coder-1.3B \cite{guo2024deepseek}, Llama-3.2-1B \cite{Llama3.2}, and Qwen2.5-1.5B \cite{qwen2025qwen25technicalreport}. {\workname} defaults to DeepSeek-Coder-1.3B unless specified.
Micro Coding is verified on 6 powerful LLMs, including the closed-source models of Gemini 2.5 Pro \cite{comanici2025gemini}, Gemini 2.5 Flash \cite{comanici2025gemini}, Claude Sonnet 4 \cite{claude4} and OpenAI o4-mini \cite{o4-mini}, and the open-source models of DeepSeek-V3 \cite{liu2024deepseek} and DeepSeek-R1 \cite{guo2025deepseek}.

\textbf{Hardware.} 
{\workname} is tested on three GPUs in Table \ref{tab:gpu_comparison} with the following CPUs: Intel Silver 4114 (V100), Intel Gold 6336Y (A100), and Intel Platinum 8575C (H100).

\begin{table}[htbp]
\centering
\scriptsize
\renewcommand{\arraystretch}{0.75}
\setlength{\tabcolsep}{7pt}
\begin{tabular}{l ccc}
    \toprule
    \textbf{Feature} & \textbf{V100} & \textbf{A100} & \textbf{H100} \\
    \midrule
    Architecture         & Volta      & Ampere     & Hopper     \\
    SMs (Streaming Multiprocessors) & 80         & 108        & 132        \\
    Global Memory (GB) & 32 & 80 & 80 \\
    Shared Memory / SM (KB)         & 96         & 164        & 228        \\
    L2 Cache (MB)                   & 6          & 40         & 50         \\
    Memory Bandwidth (GB/s)         & 900        & 1935       & 3350       \\
    FP32 TFLOPS                     & 15.7       & 19.5       & 60         \\
    \bottomrule
\end{tabular}
\caption{The main features of diverse GPU platforms.}
\label{tab:gpu_comparison}
\end{table}

\textbf{Benchmarks.}
{\workname} is evaluated on KernelBench and TritonBench, as depicted in Table \ref{tab:kb_details}.
Specifically, KernelBench is tailored to benchmark LLMs in correct and efficient GPU kernel generation across 250 tasks at three levels.
Similarly, TritonBench assesses LLM capabilities in Triton kernel generation via 184 real-world kernels (TRITONBENCH-G) and 166 PyTorch-aligned interface kernels (TRITONBENCH-T).

\textbf{Metrics.}
Following the benchmark setting and prior work, we adopt four metrics to comprehensively evaluate both correctness and performance of generated kernels.
\textbf{Call Accuracy} (only adopted in TritonBench) and \textbf{Execute Accuracy} measure the rates of kernels that successfully pass compilation and produce correct computation results, respectively.
$\textbf{fast}_p$ quantifies the proportion of kernels, which are correct and achieve speedup $\textgreater p$ versus PyTorch Eager, across all $N$ tasks. \textbf{Mean Speedup} represents the arithmetic mean of speedups. They are calculated as follows:

\begin{equation}
\small
    \text{fast}_p = \frac{1}{N}\sum^{N}_{i=1} 1 (\text{correct}_i \land  \{\text{speedup}_i > p\})
    \label{eq:fastp}
\end{equation}
\begin{equation}
\small
\text{Mean Speedup} = \frac{1}{N}\sum^{N}_{i=1} \text{speedup}_i
\label{eq:mean}
\end{equation}


\textbf{Baselines.}
MTMC undergoes comprehensive comparison against multiple baselines including expert-optimized PyTorch Eager kernels, SOTA general-purpose LLMs (see Table \ref{tab:main_kb} and \ref{tab:main_tb}), domain-specific LLMs (Qwen2.5-Coder-32B \cite{hui2024qwen2}) and agent (Gemini CLI \cite{GeminiCLI}) for code generation, and intensively finetuned LLMs for GPU kernel generation (Kevin-32B \cite{multi-turn-kernels} and KernelLLM \cite{kernelllm2025}). See Appendix for details on the baseline setup.


\subsection{Overall Performance}
Tables~\ref{tab:main_kb} and~\ref{tab:main_tb} respectively show the results of {\workname} across KernelBench and TritonBench,
which underscores that {\workname} achieves \textbf{new SOTA} that empowers LLMs to produce \textbf{both correct and high-performing} GPU kernels across diverse hardware and tasks.

KernelBench results demonstrate that {\workname} is the \textbf{only approach approaching or surpassing expert-optimized kernel performance} in PyTorch Eager (significant \textgreater 1× speedup at Levels 1-2 with near-perfect accuracy), dramatically outperforming baselines.
Specifically, \textbf{versus general-purpose LLMs and code LLMs/agent}: accuracy improves by up to 50\% (KernelBench-L2/TritonBench-T) with speedup gains up to 5× (KernelBench-L3).
Moreover, \textbf{compared to finetuned LLMs} (Kevin-32B/KernelLLM), MTMC delivers \textgreater 50\% accuracy gains across both benchmarks and achieves 32-34× performance in TritonBench.
    
\begin{table*}[htbp]
\centering
\scriptsize
\renewcommand{\arraystretch}{0.8}
\setlength{\tabcolsep}{0.8pt}
\begin{tabular}{l|l|ccc|ccc|ccc}
    \toprule
    \multirow{3}{*}{\textbf{Hardware}} & \multirow{3}{*}{\textbf{Method}} & \multicolumn{3}{c|}{Level 1} & \multicolumn{3}{c|}{Level 2} & \multicolumn{3}{c}{Level 3} 
    \\
    \cmidrule(lr){3-5}
    \cmidrule(lr){6-8}
    \cmidrule(lr){9-11}
    & & \makecell{Accuracy(\%)} & \makecell{fast$_1$/fast$_2$(\%)} & \makecell{Mean Speedup}
    & \makecell{Accuracy (\%)} & \makecell{fast$_1$/fast$_2$(\%)} & \makecell{Mean Speedup}
    & \makecell{Accuracy(\%)} & \makecell{fast$_1$/fast$_2$(\%)} & \makecell{Mean Speedup}
    \\
    \midrule
    \multirow{20}{*}{\textbf{H100}}
    & Claude-3.7-Sonnet & 32 & 10 / 2 & 0.30 & 11 & 7 / 0 & 0.11 & 12 & 2 / 0 & 0.09 \\ 
    & Claude-4-Sonnet & 50 & 26 / 6 & 1.20 & 41 & 32 / 2 & 0.49 & 20 & 6 / 0 & 0.14 \\ 
    & OpenAI o4-mini & 48 & 23 / 4 & 1.07 & 38 & 25 / 1 & 0.44 & 26 & 6 / 0 & 0.21 \\ 
    & GPT-4o & 20 & 13 / 3 & 0.50 & 2 & 2 / 0 & 0.02 & 6 & 2 / 0 & 0.06 \\ 
    & DeepSeek-R1 & 52 & 23 / 5 & 1.18 & 48 & \underline{38} / \underline{4} & 0.65 & 28 & 8 / 0 & 0.23 \\ 
    & DeepSeek-V3-0324 & 30 & 14 / 4 & 0.92 & 1 & 1 / 0 & 0.01 & 4 & 2 / 0 & 0.04 \\ 
    & Llama-3.1-Nemontron & 16 & 10 / 1 & 0.18 & 7 & 5 / 0 & 0.07 & 4 & 2 / 0 & 0.04 \\ 
    & Qwen3-253B-A22B & 49 & 18 / 4 & 0.97 & 39 & 15 / 2 & 0.33 & 10 & 2 / 0 & 0.08 \\ 
    \cmidrule(lr){2-11}
    & Qwen2.5-Coder-32B & 14 & 8 / 1 & 0.70 & 1 & 1 / 0 & 0.01 & 4 & 0 / 0 & 0.02 \\ 
    & Gemini CLI & 51 & \underline{32} / 3 & 1.06 & 32 & 25 / 3 & 0.38 & 22 & \underline{14} / 0 & 0.20 \\ 
    \cmidrule(lr){2-11}
    & Kevin-32B (Stanford)& \underline{68} & 9 / 2 & 0.71 & \underline{68} & 24 / 2 & 0.58 & \underline{48} & 4 / 0 & \underline{0.35} \\
    & KernelLLM (Meta) & 41 & 11 / 2 & 0.38 & 35 & 20 / 1 & 0.41 & 10 & 2 / 0 & 0.09 \\ 
    \cmidrule(lr){2-11}
    & Gemini 2.5 Pro & 63 & 31 / 7 & \underline{1.26} & 57 & 34 / \underline{4} & \underline{0.77} & 36 & 6 / 0 & 0.27 \\ 

    & \cellcolor{gray!20} + Ours & \cellcolor{gray!20} \textbf{100}($\uparrow$37\%) & \cellcolor{gray!20} \textbf{67/13}($\uparrow$36\%/6\%) & \cellcolor{gray!20} \textbf{2.08}($\uparrow$1.65$\times$) & \cellcolor{gray!20} \textbf{99}($\uparrow$42\%) & \cellcolor{gray!20} \textbf{86/12}($\uparrow$52\%/8\%) & \cellcolor{gray!20} \textbf{1.28}($\uparrow$1.66$\times$) & \cellcolor{gray!20} \textbf{70}($\uparrow$ 34\%) & \cellcolor{gray!20} \textbf{34/2}($\uparrow$28\%/2\%) & \cellcolor{gray!20} \textbf{0.77}($\uparrow$2.85$\times$) \\
    & \cellcolor{gray!20} v.s. Kevin-32B & \cellcolor{gray!20} $\uparrow$ 32\% & \cellcolor{gray!20} $\uparrow$ 58\% / 11\% & \cellcolor{gray!20} $\uparrow$ 2.93$\times$ & \cellcolor{gray!20} $\uparrow$ 31\% & \cellcolor{gray!20} $\uparrow$ 62\% / 10\% & \cellcolor{gray!20} $\uparrow$ 2.10$\times$ & \cellcolor{gray!20} $\uparrow$ 22\% & \cellcolor{gray!20} $\uparrow$ 30\% / 2\% & \cellcolor{gray!20} $\uparrow$ 8.41 $\times$
    \\

    & \cellcolor{gray!20} v.s. KernelLLM & \cellcolor{gray!20} $\uparrow$ 59\% & \cellcolor{gray!20} $\uparrow$ 56\% / 11\% & \cellcolor{gray!20} $\uparrow$ 5.47$\times$ & \cellcolor{gray!20} $\uparrow$ 64\% & \cellcolor{gray!20} $\uparrow$ 66\% / 11\% & \cellcolor{gray!20} $\uparrow$ 3.12$\times$ & \cellcolor{gray!20} $\uparrow$ 60\% & \cellcolor{gray!20} $\uparrow$ 32\% / 2\% & \cellcolor{gray!20} $\uparrow$ 8.56 $\times$
    \\
    
    & Gemini 2.5 Flash & 54 & 29 / \underline{9} & 1.25 & 47 & 30 / 3 & 0.53 & 32 & 4 / 0 & 0.15 \\ 
    & \cellcolor{gray!20} + Ours & \cellcolor{gray!20} \textbf{94}($\uparrow$40\%) & \cellcolor{gray!20} \textbf{54/13}($\uparrow$ 25\%/4\%) & \cellcolor{gray!20} \textbf{2.03}($\uparrow$1.62$\times$) & \cellcolor{gray!20} \textbf{97}($\uparrow$50\%) & \cellcolor{gray!20} \textbf{85/5}($\uparrow$55\%/2\%) & \cellcolor{gray!20} \textbf{1.35}($\uparrow$2.55$\times$) & \cellcolor{gray!20} \textbf{64}($\uparrow$32\%) & \cellcolor{gray!20} \textbf{28/2}($\uparrow$ 24\%/2\%) & \cellcolor{gray!20} \textbf{0.73}($\uparrow$4.87$\times$) \\ 
    & \cellcolor{gray!20} v.s. Kevin-32B & \cellcolor{gray!20} $\uparrow$ 26\% & \cellcolor{gray!20} $\uparrow$ 45\% / 11\% & \cellcolor{gray!20} $\uparrow$ 2.86$\times$ & \cellcolor{gray!20} $\uparrow$ 29\% & \cellcolor{gray!20} $\uparrow$ 61\% / 3\% & \cellcolor{gray!20} $\uparrow$ 2.33$\times$ & \cellcolor{gray!20} $\uparrow$ 16\% & \cellcolor{gray!20} $\uparrow$ 24\% / 2\% & \cellcolor{gray!20} $\uparrow$  2.09$\times$
    \\

    & \cellcolor{gray!20} v.s. KernelLLM & \cellcolor{gray!20} $\uparrow$ 53\% & \cellcolor{gray!20} $\uparrow$ 43\% / 11\% & \cellcolor{gray!20} $\uparrow$ 5.34$\times$ & \cellcolor{gray!20} $\uparrow$ 62\% & \cellcolor{gray!20} $\uparrow$ 65\% / 4\% & \cellcolor{gray!20} $\uparrow$ 3.29$\times$ & \cellcolor{gray!20} $\uparrow$ 54\% & \cellcolor{gray!20} $\uparrow$ 26\% / 2\% & \cellcolor{gray!20} $\uparrow$  8.11$\times$
    \\
    \midrule
    \multirow{20}{*}{\textbf{A100}} 
    & Claude-3.7-Sonnet & 32 & 8 / 3 & 0.32 & 11 & 4 / 0 & 0.10 & 12 & 6 / 0 & 0.10 \\ 
    & Claude-4-Sonnet & 50 & 15 / 5 & 1.31 & 41 & 15 / 1 & 0.39 & 20 & 6 / 0 & 0.15 \\ 
    & OpenAI o4-mini & 48 & 13 / 3 & 1.27 & 38 & 25 / 0 & 0.39 & 28 & 8 / \underline{2} & 0.26 \\ 
    & GPT-4o & 20 & 8 / 2 & 0.57 & 2 & 2 / 0 & 0.02 & 6 & 6 / 0 & 0.06 \\ 
    & DeepSeek-R1 & 52 & 15 / 4 & 1.37 & 46 & 33 / 1 & 0.50 & 28 & 12 / 0 & 0.27 \\ 
    & DeepSeek-V3-0324 & 30 & 6 / 3 & 1.12 & 1 & 0 / 0 & 0.01 & 4 & 2 / 0 & 0.04 \\ 
    & Llama-3.1-Nemontron & 16 & 5 / 0 & 0.16 & 7 & 5 / 0 & 0.07 & 4 & 4 / 0 & 0.04 \\ 
    & Qwen3-253B-A22B & 49 & 8 / 3 & 1.18 & 38 & 12 / \underline{2} & 0.27 & 10 & 2 / 0 & 0.08 \\ 
    \cmidrule(lr){2-11}
    & Qwen2.5-Coder-32B & 14 & 3 / 1 & 0.94 & 1 & 1 / 0 & 0.01 & 4 & 2 / 0 & 0.02 \\ 
    & Gemini CLI & 51 & \underline{30} / 2 & 1.29 & 32 & 22 / 1 & 0.34 & 22 & \underline{16} / 0 & 0.20 \\ 
    \cmidrule(lr){2-11}
    & Kevin-32B (Stanford)& \underline{69} & 4 / 3 & 0.84 & \underline{67} & 8 / 0 & 0.40 & \underline{46} & 2 / 0 & \underline{0.32} \\ 
    & KernelLLM (Meta)& 41 & 21 / 2 & 0.50 & 35 & \underline{34} / \underline{2} & 0.50 & 10 & 6 / 0 & 0.10 \\ 
    \cmidrule(lr){2-11}
    & Gemini 2.5 Pro & 63 & 23 / 6 & \underline{1.47} & 57 & 33 / \underline{2} & \underline{0.69} & 36 & 8 / \underline{2} & 0.14 \\



    & \cellcolor{gray!20} + Ours & \cellcolor{gray!20} \textbf{100}($\uparrow$37\%) & \cellcolor{gray!20} \textbf{59/9}($\uparrow$36\%/3\%) & \cellcolor{gray!20} \textbf{2.20}($\uparrow$1.50$\times$) & \cellcolor{gray!20} \textbf{99}($\uparrow$42\%) & \cellcolor{gray!20} \textbf{66/8}($\uparrow$33\%/6\%) & \cellcolor{gray!20} \textbf{1.22}($\uparrow$1.77$\times$) & \cellcolor{gray!20} \textbf{70}($\uparrow$34\%) & \cellcolor{gray!20} \textbf{20/4}($\uparrow$12\%/2\%) & \cellcolor{gray!20} \textbf{0.73}($\uparrow$5.21$\times$) \\

    & \cellcolor{gray!20} v.s. Kevin-32B & \cellcolor{gray!20} $\uparrow$ 31\% & \cellcolor{gray!20} $\uparrow$ 55\% / 6\% & \cellcolor{gray!20} $\uparrow$ 2.62$\times$ & \cellcolor{gray!20} $\uparrow$ 32\% & \cellcolor{gray!20} $\uparrow$ 58\% / 8\% & \cellcolor{gray!20} $\uparrow$ 3.05$\times$ & \cellcolor{gray!20} $\uparrow$ 24\% & \cellcolor{gray!20} $\uparrow$ 18\% / 4\% & \cellcolor{gray!20} $\uparrow$  2.28$\times$
    \\

    & \cellcolor{gray!20} v.s. KernelLLM & \cellcolor{gray!20} $\uparrow$ 59\% & \cellcolor{gray!20} $\uparrow$ 38\% / 7\% & \cellcolor{gray!20} $\uparrow$ 4.40$\times$ & \cellcolor{gray!20} $\uparrow$ 64\% & \cellcolor{gray!20} $\uparrow$ 32\% / 6\% & \cellcolor{gray!20} $\uparrow$ 2.44$\times$ & \cellcolor{gray!20} $\uparrow$ 60\% & \cellcolor{gray!20} $\uparrow$ 14\% / 4\% & \cellcolor{gray!20} $\uparrow$  7.30$\times$
    \\
    & Gemini 2.5 Flash & 54 & 21 / \underline{7} & 1.42 & 47 & 30 / \underline{2} & 0.51 & 32 & 4 / 0 & 0.12 \\ 
    
    & \cellcolor{gray!20} + Ours & \cellcolor{gray!20} \textbf{94}($\uparrow$40\%) & \cellcolor{gray!20} \textbf{53/10}($\uparrow$32\%/3\%) & \cellcolor{gray!20} \textbf{2.14}($\uparrow$1.51$\times$) & \cellcolor{gray!20} \textbf{97}($\uparrow$50\%) & \cellcolor{gray!20} \textbf{56/4}($\uparrow$26\%/2\%) & \cellcolor{gray!20} \textbf{1.21}($\uparrow$2.37$\times$) & \cellcolor{gray!20} \textbf{64}($\uparrow$32\%) & \cellcolor{gray!20} \textbf{48/2}($\uparrow$44\%/2\%) & \cellcolor{gray!20} \textbf{0.69}($\uparrow$5.75$\times$) \\ 
    
    & \cellcolor{gray!20} v.s. Kevin-32B & \cellcolor{gray!20} $\uparrow$ 25\% & \cellcolor{gray!20} $\uparrow$ 49\% / 7\% & \cellcolor{gray!20} $\uparrow$ 2.55$\times$ & \cellcolor{gray!20} $\uparrow$ 30\% & \cellcolor{gray!20} $\uparrow$ 48\% / 4\% & \cellcolor{gray!20} $\uparrow$ 3.03$\times$ & \cellcolor{gray!20} $\uparrow$ 18\% & \cellcolor{gray!20} $\uparrow$ 46\% / 2\% & \cellcolor{gray!20} $\uparrow$  2.16$\times$
    \\

    & \cellcolor{gray!20} v.s. KernelLLM & \cellcolor{gray!20} $\uparrow$ 53\% & \cellcolor{gray!20} $\uparrow$ 32\% / 8\% & \cellcolor{gray!20} $\uparrow$ 4.28$\times$ & \cellcolor{gray!20} $\uparrow$ 62\% & \cellcolor{gray!20} $\uparrow$ 22\% / 2\% & \cellcolor{gray!20} $\uparrow$ 2.42$\times$ & \cellcolor{gray!20} $\uparrow$ 54\% & \cellcolor{gray!20} $\uparrow$ 42\% / 2\% & \cellcolor{gray!20} $\uparrow$  6.90$\times$
    \\
    \midrule
    \multirow{20}{*}{\textbf{V100}} 
    & Claude-3.7-Sonnet & 32 & 7 / 2 & 0.33 & 11 & 2 / 0 & 0.10 & 6 & 4 / 0 & 0.06 \\ 
    & Claude-4-Sonnet & 50 & \underline{17} / 5 & 1.17 & 41 & 29 / 0 & 0.46 & 20 & 0 / 0 & 0.14 \\ 
    & OpenAI o4-mini & 48 & 15 / 3 & 1.06 & 38 & 25 / 1 & 0.40 & 26 & 2 / 0 & 0.22 \\ 
    & GPT-4o & 20 & 9 / 2 & 0.63 & 2 & 0 / 0 & 0.02 & 6 & 0 / 0 & 0.06 \\ 
    & DeepSeek-R1 & 51 & 14 / 4 & 1.13 & 48 & \underline{33} / \underline{3} & 0.56 & 26 & \underline{10} / 0 & 0.24 \\ 
    & DeepSeek-V3-0324 & 30 & 7 / 3 & 0.93 & 1 & 0 / 0 & 0.01 & 4 & 2 / 0 & 0.04 \\ 
    & Llama-3.1-Nemontron & 16 & 6 / 0 & 0.17 & 7 & 5 / 0 & 0.07 & 4 & 2 / 0 & 0.04 \\ 
    & Qwen3-253B-A22B & 48 & 10 / 3 & 1.02 & 38 & 12 / \underline{3} & 0.28 & 10 & 0 / 0 & 0.09 \\ 
    \cmidrule(lr){2-11}
    & Qwen2.5-Coder-32B & 14 & 4 / 1 & 0.68 & 1 & 0 / 0 & 0.01 & 2 & 0 / 0 & 0.02 \\ 
    & Gemini CLI & 49 & 16 / 3 & 1.07 & 32 & 22 / 1 & 0.35 & 22 & 8 / 0 & 0.20 \\ 
    \cmidrule(lr){2-11}
    & Kevin-32B (Stanford) & \underline{69} & 6 / 2 & 0.86 & \underline{67} & 14 / 0 & 0.48 & \underline{46} & 4 / 0 & \underline{0.35} \\ 
    & KernelLLM (Meta) & 41 & 14 / 2 & 0.52 & 35 & 31 / 2 & 0.49 & 10 & 4 / 0 & 0.10 \\ 
    \cmidrule(lr){2-11}
    & Gemini 2.5 Pro & 63 & \underline{17} / 6 & \underline{1.27} & 57 & 29 / 2 & \underline{0.65} & 36 & 8 / 0 & 0.27 \\
    & \cellcolor{gray!20} + Ours & \cellcolor{gray!20} \textbf{99}($\uparrow$36\%) & \cellcolor{gray!20} \textbf{43/9}($\uparrow$26\%/3\%) & \cellcolor{gray!20} \textbf{1.94}($\uparrow$1.53$\times$) & \cellcolor{gray!20} \textbf{100}($\uparrow$43\%) & \cellcolor{gray!20} \textbf{51/3}($\uparrow$22\%/1\%) & \cellcolor{gray!20} \textbf{1.04}($\uparrow$1.60$\times$) & \cellcolor{gray!20} \textbf{70}($\uparrow$34\%) & \cellcolor{gray!20} \textbf{24/2}($\uparrow$16\%/2\%) & \cellcolor{gray!20} \textbf{0.65}($\uparrow$2.41$\times$) 
    \\


    & \cellcolor{gray!20} v.s. Kevin-32B & \cellcolor{gray!20} $\uparrow$ 30\% & \cellcolor{gray!20} $\uparrow$ 37\% / 7\% & \cellcolor{gray!20} $\uparrow$ 2.26$\times$ & \cellcolor{gray!20} $\uparrow$ 33\% & \cellcolor{gray!20} $\uparrow$ 37\% / 3\% & \cellcolor{gray!20} $\uparrow$ 2.17$\times$ & \cellcolor{gray!20} $\uparrow$ 24\% & \cellcolor{gray!20} $\uparrow$ 20\% / 2\% & \cellcolor{gray!20} $\uparrow$  1.86$\times$
    \\

    & \cellcolor{gray!20} v.s. KernelLLM & \cellcolor{gray!20} $\uparrow$ 58\% & \cellcolor{gray!20} $\uparrow$ 29\% / 7\% & \cellcolor{gray!20} $\uparrow$ 3.73$\times$ & \cellcolor{gray!20} $\uparrow$ 65\% & \cellcolor{gray!20} $\uparrow$ 20\% / 1\% & \cellcolor{gray!20} $\uparrow$ 2.12$\times$ & \cellcolor{gray!20} $\uparrow$ 60\% & \cellcolor{gray!20} $\uparrow$ 20\% / 2\% & \cellcolor{gray!20} $\uparrow$  6.50$\times$
    \\

    & Gemini 2.5 Flash & 53 & 16 / \underline{8} & 1.15 & 47 & 27 / 1 & 0.45 & 32 & 4 / 0 & 0.15 \\ 

    & \cellcolor{gray!20} + Ours & \cellcolor{gray!20} \textbf{93}($\uparrow$40\%) & \cellcolor{gray!20} \textbf{45/11}($\uparrow$29\%/3\%) & \cellcolor{gray!20} \textbf{1.93}($\uparrow$1.68$\times$) & \cellcolor{gray!20} \textbf{97}($\uparrow$50\%) & \cellcolor{gray!20} \textbf{56/3}($\uparrow$29\%/2\%) & \cellcolor{gray!20} \textbf{1.18}($\uparrow$2.62$\times$) & \cellcolor{gray!20} \textbf{64}($\uparrow$32\%) & \cellcolor{gray!20} \textbf{36/2}($\uparrow$32\%/2\%) & \cellcolor{gray!20} \textbf{0.68}($\uparrow$4.53$\times$) 
    \\ 

    & \cellcolor{gray!20} v.s. Kevin-32B & \cellcolor{gray!20} $\uparrow$ 24\% & \cellcolor{gray!20} $\uparrow$ 39\% / 9\% & \cellcolor{gray!20} $\uparrow$ 2.24$\times$ & \cellcolor{gray!20} $\uparrow$ 30\% & \cellcolor{gray!20} $\uparrow$ 42\% / 3\% & \cellcolor{gray!20} $\uparrow$ 2.46$\times$ & \cellcolor{gray!20} $\uparrow$ 18\% & \cellcolor{gray!20} $\uparrow$ 32\% / 2\% & \cellcolor{gray!20} $\uparrow$  1.94$\times$
    \\
    
    & \cellcolor{gray!20} v.s. KernelLLM & \cellcolor{gray!20} $\uparrow$ 52\% & \cellcolor{gray!20} $\uparrow$ 31\% / 9\% & \cellcolor{gray!20} $\uparrow$ 3.71$\times$ & \cellcolor{gray!20} $\uparrow$ 62\% & \cellcolor{gray!20} $\uparrow$ 25\% / 1\% & \cellcolor{gray!20} $\uparrow$ 2.41$\times$ & \cellcolor{gray!20} $\uparrow$ 54\% & \cellcolor{gray!20} $\uparrow$ 32\% / 2\% & \cellcolor{gray!20} $\uparrow$  6.80$\times$
    \\
    \bottomrule
\end{tabular}

\caption{
The execute accuracy and mean speedup relative to PyTorch Eager for Triton kernel generation on KernelBench across various hardware, LLMs and agent. The best results are highlighted in \textbf{bold}, while the second-best results are \underline{underlined}.}
\label{tab:main_kb}
\end{table*}

\begin{table*}[t]
\centering
\scriptsize
\setlength{\tabcolsep}{1pt}
\renewcommand{\arraystretch}{0.7}
\begin{tabular}{l | cccc | cccc}
    \toprule
    \multirow{3}{*}{\textbf{Method}} & \multicolumn{4}{c|}{TRITONBENCH-G} & \multicolumn{4}{c}{TRITONBENCH-T} \\ 
    \cmidrule(lr){2-5} \cmidrule(lr){6-9}
    & \makecell{Call Accuracy(\%)} & \makecell{Execute Accuracy(\%)}  & \makecell{fast$_1$/fast$_2$(\%)} & \makecell{Mean Speedup}
    & \makecell{Call Accuracy(\%)} & \makecell{Execute Accuracy(\%)}  & \makecell{fast$_1$/fast$_2$(\%)} & \makecell{Mean Speedup} \\
    \midrule
        Gemini 2.5 Pro & 25.00 & 16.00 & \underline{9.24} / 1.09 & \underline{0.28} & \underline{28.92} & \underline{22.89} & \underline{6.63} / \underline{1.81} & \underline{0.24} \\ 
        
        Claude-3.7-Sonnet & 11.41 & 8.70 & 3.80 / 0.54 & 0.17 & 14.46 & 9.64 & 2.41 / 0.60 & 0.08 \\ 
        Claude-4-Sonnet & \underline{25.54} & \underline{17.39} & 7.07 / 0.54 & 0.27 & 16.27 & 10.84 & 3.01 / 1.20 & 0.10 \\ 
        OpenAI o4-mini & 11.96 & 7.07 & 3.26 / 1.09 & 0.18 & 8.43 & 7.83 & 1.81 / 0.60 & 0.09 \\ 
        GPT-4o & 5.98 & 3.80 & 1.63 / 0.54 & 0.13 & 6.63 & 2.41 & 1.20 / 0.00 & 0.03 \\ 
        DeepSeek-R1-0528 & 15.76 & 8.70 & 5.98 / 1.09 & 0.23 & 28.31 & 17.47 & 6.02 / 1.20 & 0.20 \\
        DeepSeek-V3-0324 &  11.41 & 7.61 & 3.26 / 0.54 & 0.17 & 18.67 & 13.86 & 4.82 / 0.60 & 0.15 \\ 
        \midrule
        Qwen2.5-Coder-32B & 4.35 & 3.26 & 1.09 / 1.09 & 0.08 & 4.82 & 3.01 & 0.60 / 0.00 & 0.03 \\
        \midrule
        KernelLLM (Meta)& 2.17 & 1.09 & 0.54 / 0.00 & 0.01 & 4.82 & 4.22 & 0.60 / 0.00 & 0.02 \\
        \midrule

        Gemini 2.5 Flash & 11.41 & 8.70 & 4.35 / \underline{1.63} & 0.24 & 14.46 & 9.04 & 5.42 / \underline{1.81} & 0.15 \\ 
        
        \cellcolor{gray!20} + Ours & \cellcolor{gray!20} \textbf{32.61} & \cellcolor{gray!20} \textbf{22.83} & \cellcolor{gray!20} \textbf{9.78 / 1.63} & \cellcolor{gray!20} \textbf{0.34} & \cellcolor{gray!20} \textbf{64.46} & \cellcolor{gray!20} \textbf{54.82} & \cellcolor{gray!20} \textbf{19.28 / 3.01} & \cellcolor{gray!20} \textbf{0.64} \\
        
       \cellcolor{gray!20} & \cellcolor{gray!20} $\uparrow$ 21.2\% & \cellcolor{gray!20} $\uparrow$ 14.13\% & \cellcolor{gray!20} $\uparrow$ 5.41\% / - & \cellcolor{gray!20} $\uparrow$ 1.42$\times$ & \cellcolor{gray!20} $\uparrow$ 50.00\% & \cellcolor{gray!20} $\uparrow$ 45.78\% & \cellcolor{gray!20} $\uparrow$ 13.86\% / 1.20\% & \cellcolor{gray!20} $\uparrow$ 4.67$\times$  \\


       \cellcolor{gray!20} v.s. KernelLLM & \cellcolor{gray!20} $\uparrow$ 30.44\% & \cellcolor{gray!20} $\uparrow$ 21.74\% & \cellcolor{gray!20} $\uparrow$ 9.24\% / 1.63\% & \cellcolor{gray!20} $\uparrow$ 34.00$\times$ & \cellcolor{gray!20} $\uparrow$ 59.64\% & \cellcolor{gray!20} $\uparrow$ 50.60\% & \cellcolor{gray!20} $\uparrow$ 19.22\% / 3.01\% & \cellcolor{gray!20} $\uparrow$ 32.00$\times$
       \\
    \bottomrule
\end{tabular}
\caption{
Performance comparison across LLMs and agents for Triton kernel generation on TritonBench with A100 GPU.
}
\label{tab:main_tb}
\end{table*}

\textbf{{\workname} achieves co-optimization of correctness and performance.}
Defying conventional correctness-performance trade-offs in high-performance kernel generation, MTMC’s decoupled generation strategy achieves simultaneous gains. 
On KernelBench, it delivers dramatic accuracy increase (near 100\% accuracy on levels 1 and 2) and up to several-fold improvement in mean speedup (level 3).
In contrast, vanilla LLMs generally underperform, while the finetuned LLM (Kevin-32B) significantly sacrifices performance in favor of correctness.
On TritonBench, we also achieve SOTA performance and demonstrate substantial gains over the baselines.

\textbf{{\workname} demonstrates strong generalization capabilities across diverse benchmarks and hardware platforms.}
Despite distinct task characteristics, {\workname}~achieves significant improvements in correctness and performance over general-purpose LLMs on both benchmarks.
Notably, the finetuned LLM (KernelLLM) exhibits severe degradation from KernelBench to TritonBench, accuracy from 40-50\% to 2-4\% and speedup from 0.5$\times$ to 0.01$\times$, even performing significantly worse than general-purpose LLMs. This confirms the effectiveness of our decoupled, layered generation mechanism. Furthermore, KernelBench results show our generated kernels deliver consistent performance gains across diverse GPU architectures (V100 to A100, spanning three generations). This suggests our Macro Thinking policy learns universal optimization strategies from limited data.

\textbf{{\workname} ~maintains robust performance across tasks of varying difficulty levels.}
MTMC achieves SOTA on varying-difficulty tasks across both benchmarks, significantly surpassing baselines. 
In contrast, general-purpose/code/finetuned LLMs all show inconsistent results (Tables 1-2, underlined suboptimal entries), due to their exclusive reliance on LLMs' semantic knowledge, causing substantial randomness.
{\workname}'s decoupled design enables robust performance: Macro Thinking learns optimization strategies while Micro Coding generates implementations.

\subsection{Ablation Study}

\begin{table}[htbp]
\centering
\scriptsize
\setlength{\tabcolsep}{6pt}
\renewcommand{\arraystretch}{0.4}
\begin{tabular}{l ccccccc}
\toprule
    \makecell[l]{Task ID} & 1 & 2  & 6 & 7 & 8 & 9 & 13 \\
    \midrule
    \makecell[l]{{\workname} (Triton)} & 1.38 & 1.36 & 4.43 & 1.45 & 37.88 & 0.92 & 10.56 \\
    \makecell[l]{{\workname} (CUDA)} & 1.38 & 1.66 & 1.34 & 1.66 & 26.52 & 0.91 & 10.17 \\
\bottomrule
\end{tabular}
\caption{Execution time (ms
) of {\workname} on KernelBench \texttt{matmul} operators with different generation targets.}
\label{tab:ab_target}
\end{table}

\textbf{Ablation of Target Programming Language: 
MTMC demonstrates scalability to additional kernel programming languages.}
While our experiments primarily target Triton due to LLMs' difficulties with low-level CUDA generation (sparse corpus and high complexity), Table \ref{tab:ab_target} reveals MTMC can produce high-performance CUDA kernels with LLMs' greater familiarity operators (like \texttt{matmul}).
Thus, the primary scalability bottleneck resides in the LLM's proficiency with the target programming language.

\textbf{Ablation of Micro Coding: Hierarchical kernel generation with multi-step implementation is critical.} 
As shown in Table \ref{tab:ab_decouple}, feeding all optimization actions and corresponding prompts at once to the LLM (w/o Hier) causes significant performance degradation (up to 64\% accuracy and 1.3$\times$ speedup). This indicates that current SOTA LLMs cannot generate such complex kernels in a single pass, confirming the rationale behind our hierarchical step-by-step optimization and implementation design.
  
  
\begin{table}[htbp]
\centering
\scriptsize
\renewcommand{\arraystretch}{0.8}
\setlength{\tabcolsep}{3pt}
\begin{tabular}{l ccc}
    \toprule
     KernelBench Level & Level 1 & Level 2 & Level 3 \\
    \cmidrule{2-2} \cmidrule{3-3} \cmidrule{4-4}
    & \makecell{Accuracy/Speedup} & \makecell{Accuracy/Speedup} & \makecell{Accuracy/Speedup} \\
    \midrule
    GF-2.5 w/o Hier & 60\% / 1.38 & 32\% / 0.43 & 10\% / 0.09 \\
    GF-2.5 + Ours & \textbf{94\%} / \textbf{2.14} & \textbf{97\%} / \textbf{1.21} & \textbf{64\%} / \textbf{0.69} \\
    \midrule
    DS-V3 w/o Hier  & 41\% / 0.52 & 16\% / 0.17 & 6\% / 0.04 \\
    DS-V3 + Ours & \textbf{78\%} / \textbf{1.82} & \textbf{59\%} / \textbf{0.66} & \textbf{36\%} / \textbf{0.32} \\
    \bottomrule
\end{tabular}
\caption{Comparison between multi-step (ours) and single-pass generation (w/o Hier).}
\label{tab:ab_decouple}
\end{table}
\textbf{Ablation of Macro Thinking:}
Table \ref{tab:ab_agent} presents ablation on two key designs of the Macro Thinking component:``policy'' indicating whether to learn optimization policies, and ``AS'' determining whether to leverage action spaces based on hardware-aware optimization summary.
\textbf{(1) The efficacy of the Macro Thinking policy exhibits robustness to the choice of lightweight LLMs.} Policies with diverse LLMs consistently achieve high correctness and speedup, with the smallest model paradoxically delivering the best results. This demonstrates the exceptional efficiency of our policy training paradigm.
\textbf{(2) Reinforcement learning for optimization policies is essential}. Directly employing LLMs for Macro Thinking results in a marked performance decrease, despite task simplification through decoupling.
\textbf{(3) The action space design proves both rational and critical.} Without this, unrestricted LLM-generated suggestions cause further performance degradation. This finding confirms LLMs' fundamental lack of comprehension regarding hardware-aware optimizations.


\begin{table}[htbp]
\centering
\scriptsize
\setlength{\tabcolsep}{6pt}
\renewcommand{\arraystretch}{0.8}
\begin{tabular}{l l ccc}
  \toprule
  \multicolumn{1}{l}{Setting} & Method & Level 1 & Level 2 & Level 3 \\
  \midrule

  \multirow{3}{*}{\makecell{w/ policy \\ w/ AS}} 
    & - DS-Coder  & 90\% / 1.10 & 100\% / 1.16 & 100\% / 1.82 \\
    & - Llama     & 100\% / 1.17 & 80\% / 0.86 & 80\% / 0.74 \\ 
    & - Qwen      & 90\% / 0.97 & 100\% / 0.90 & 100\% / 0.75 \\ 
  \midrule
  \multirow{4}{*}{\makecell{w/o policy \\ w/ AS}} 
    & - random & 70\% / 0.50 & 50\% / 0.51 & 40\% / 0.15 \\
    & - GPT-4o & 50\% / 0.74 & 40\% / 0.51 & 40\% / 0.29 \\ 
    & - DS-V3  & 50\% / 0.67 & 50\% / 0.60 & 60\% / 0.61 \\ 
    & - GF-2.5 & 50\% / 0.82 & 60\% / 0.62 & 60\% / 0.53 \\
  \midrule
  \multirow{3}{*}{\makecell{w/o policy \\ w/o AS}}
    & - GPT-4o & 20\% / 0.16 & 30\% / 0.28 & 20\% / 0.03 \\
    & - DS-V3  & 30\% / 0.14 & 10\% / 0.02 & 20\% / 0.34 \\
    & - GF-2.5 & 30\% / 0.25 & 50\% / 0.49 & 40\% / 0.36 \\
\bottomrule
  
\end{tabular}
\caption{
The execute accuracy and mean speedup comparison of different policies and settings. 10\% of KernelBench tasks are used for testing.
}
\label{tab:ab_agent}
\end{table}





There are also two ablations in the Appendix: one on Micro Coding models that further demonstrates \workname's generalization across LLMs, and another indicating that \workname{} reaches peak performance with only a few optimization steps, while LLMs cannot promote through resampling.

\section{Conclusion}
In this paper, we propose {\workname}, a hierarchical framework that decouples high-level optimization strategy and low-level implementation and enables LLMs to automatically generate correct and high-performance GPU kernels.

Results on widely adopted benchmarks show {\workname} significantly outperforms existing LLMs in both SOTA correctness and performance, and is the only method significantly surpassing expert-optimized PyTorch Eager kernels in most tasks.
MTMC currently focuses on GPU / Triton and can be further improved in the network-level kernel (already significantly surpassing other works). We will extend it to more emerging hardware platforms and languages, and further promote its ability in future work.

\section{Acknowledgments}
This work is partially supported by the NSF of China (under Grant 92364202), and Major Program of ISCAS (Grant No. ISCAS-ZD-202402).

\bibliography{aaai2026}

\end{document}